# Système de représentation d'aide au besoin dans le domaine architectural

Marie-France ANGO -OBIANG

**Doctorante**
ango@loria.fr

Equipe SITE, Laboratoire Lorrain de Recherche en informatique et ses Applications (LORIA)

Campus Scientifique BP 329, 54506 Vandoeuvre-lès-Nancy. France. Tél : + 33 3 83 59 20 87

Fax : + 33 3 83 27 83 19

**Résumé**

L'image est un moyen de communication très important dans le domaine de l'architecture qui intervient dans les différentes phases de la conception d'un projet. Elle peut être considérée comme un outil d'aide à la décision. L'étude de notre recherche a pour objectif de voir l'apport de l'Intelligence Economique dans la résolution d'un problème décisionnel des différents partenaires (Architecte, Entrepreneur, Client) dans le domaine architectural, en vue de prendre des décisions stratégiques dans le cadre de la réalisation ou de la conception d'un ouvrage architectural. L'IE permet la prise en compte des besoins réels des utilisateurs-décideurs, de telle sorte que leurs attentes soient considérées à la première étape d'une recherche d'information et non dans l'étape finale de l'élaboration de l'outil dans l'évaluation de ce dernier.

## 1. Introduction

Les projets de construction dans le bâtiment nécessitent l'intervention de nombreux corps de métiers faisant appel à différentes compétences et expertises. Chaque corps de métier évolue indépendamment,use de terminologies et de technologies propres, et met en place des moyens spécifiques pour l'expression et la communication de l'information. Cette complexité rend difficile l'échange de données techniques basées sur des protocoles spécifiques à un domaine ou à une profession, et handicape la coopération entre les acteurs du bâtiment et l'organisation du projet. La conception architecturale est une activité surtout conceptuelle qui nécessite des savoirs multiples: esthétiques, graphiques, financiers, etc., partagés avec et entre les différents partenaires [Kacher 2005], [Bouattour 2005]. L'oeuvre architecturale est le produit d'un processus complexe où entre intuition et sensibilité individuelle, connaissances de techniques appropriées, culture, savoir-faire de l'architecte doit faire advenir l'idée et rendre celle-ci partageable, donc communicable aux différents intervenants (dans notre cas : Architecte, Entrepreneur, Client) du projet. Dans le contexte d'Intelligence Economique, l'information est prise en grande partie comme étant un facteur clé qui doit être identifié et interprété au bon moment pour aider l'utilisateur-décideur dans son processus décisionnel [Martre 1994]. Les besoins en information dans le domaine de l'architecture sont généralement mal abordés car ils génèrent des réflexions importantes sur certains aspects et dans le même temps, occultent des disfonctionnements entiers de réflexion sur les besoins et l'usage de l'information. Les outils informatiques mis en place sont parfois non utilisables, du fait qu'ils mettent l'accent sur le traitement des données et non sur la portée de la pertinence de l'action engagée et sur l'analyse du besoin informationnel réel des utilisateurs-décideurs. C'est pourquoi nous réfléchissons sur une méthodologie permettant de répondre aux besoins informationnels des potentiels décideurs-utilisateurs dans le domaine architectural. Notre étude consiste à l'analyse des stratégies de production d'informations dans le domaine architecturale avec une approche en termes d'IE [Ango-Obiang 2006].

Elle s'inscrit au croisement de deux éléments :la production d'informations comme support de la prise de décision et la mise en situation des acteurs dans les relations qu'elle entretient avec leur environnement. La question clé est d'identification, l'explicitation, voire la co-construction du besoin d'information en Architecture. De ce fait, il semble important d'effectué une analyse des enjeux des acteurs,qui nous permettra de déterminer les informations dont ils ont besoin; afin d'élaborer leurs stratégie ou, préparer les décisions qu'ils doivent prendre dans le but de se situer de manière efficace dans leur environnement.

## 2. Les acteurs dans le domaine de l' Architecture

### 2.1 Les besoins des acteurs

Répondre aux besoins exprimés d'un client c'est tout simplement lui fournir ce qu'il veut (la satisfaction du client). En satisfaisant ses exigences formalisées dans un contrat ou un cahier des charges, le fournisseur adopte une démarche qualité. Les besoins implicites, [Bourdichon 1994] quant à eux sont tout simplement l'application des règles de l'art et des normes caractérisées dans la qualité d'un produit ou d'un services rendus; Implicitement, lorsqu'un client veut l'acquisition d'une maison,la sélection d'effectue parmi différents entrepreneurs de bâtiment ,etc. En fonction de son besoin, un usager se retourne naturellement vers celui qui est reconnu comme apte à la réaliser. Pour répondre aux besoins et attentes des usagers, les entreprises doivent, plus qu'à l'écoute, anticiper sur les évolutions du marché voire même les susciter par des opérations marketing judicieusement ciblées [Bouattour 2005]. Dans ce contexte, l'aspect communication prend toute sa mesure: l'information juste doit circuler rapidement au bon moment, à l'intérieur comme à l'extérieur de l'entreprise. Par ailleurs, il faut être sûr que le partenaire possède l'information correcte et s'assurer qu'après une transmission, l'information soit bien reçue et bien perçue. Cette importance accrue de l'information comme de sa circulation se trouve être encore renforcée par les besoins plus larges du marché. Les clients ne se suffisent plus de la seule possession du produit livré. Les exigences contractuellement exprimées dans la spécification de besoins résultent d'une consultation demandeur/concepteur, démarche consensuelle permet tant d'on tenir le meilleur compromis entre le besoin réel, le coût et le délais.

### 2.2 Interaction entre les acteurs dans le bâtiment

Dans un schéma habituel, un client vient soumettre à l'architecte par exemple une intention d'habiter une maison. L'objet prend forme très progressivement et graphiquement par le savoir et savoir-faire de l'architecte. Les discussions et négociations avec le client vont permettre à l'architecte d'affiner sa proposition en fonction de ce qu'il perçoit de la demande de son interlocuteur (Figure 1).

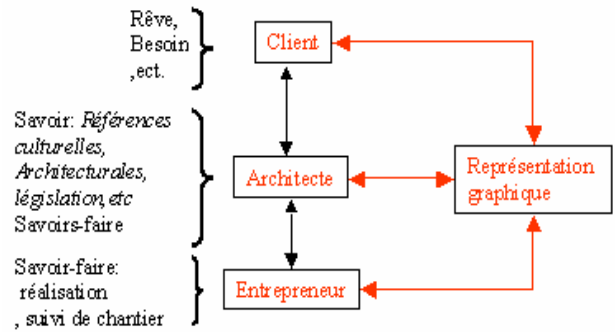

**Figure 1** : Interaction habituelle client/architecte/Entrepreneur

Nous avons retenus trois acteurs [Ango-obiang 2006] travaillant et intervenant par dialogue et/ou graphisme sur une même information visuelle qu'est l'image apporté par le client. Cette image du client est alors l'image référant de la consultation, image technique ou opérative [Lebahar 1983], il est aussi la trace visible et exploitable du passage de l'idée à la forme (dans la conception du client) ou de la forme à l'idée (dans l'explication qu'en fait le client à l'architecte). Il est surtout le support de la communication entre nos acteurs. Cette représentation sur laquelle les acteurs discutent et agissent n'a en fait aucune existence réalisée. En ce sens c'est un objet virtuel mais central, objet d'interaction où l'action est davantage orientée vers le partage, l'échange, l'interaction, la compréhension. En début de consultation, l'architecte se renseigne sur l'environnement du projet, le site,le terrain, l'orientation,le budget, etc. Les réponses sont notées comme faisant parties des contraintes du projet. Très vite un véritable dialogue s'instaure entre les acteurs pour comprendre les logiques de chacun: l'architecte pour évaluer les « contraintes » de son client et le client pour saisir les modifications, rectifications qu'il doit apporter à son projet. L'architecte doit créer et rendre réalisable un édifice.

## 3. Le vecteur de la communication

L'acteur d'une activité constitue une image mentale de son environnement,des actions et des opérations à mettre en œuvre afin de réaliser son activité. Au cours de l'action, un acteur ne montre pas dans sa représentation mentale toute la complexité de l'objet ni toutes ses propriétés.

L'acteur sélectionne l'information pertinente pour les actions qu'il souhaite mener sur l'objet. Au cours d'un travail coopératif, une orientation commune des sujets est nécessaire. Cependant, pour que l'accomplissement des actions puisse être coordonné, les orientations de tous les acteurs doivent être compatibles. Dans ce cas, le sujet doit former à la fois ses propres images opératives et se représenter les opérations, l'état des objets qu'il produit mais aussi les opérations et les objets produits par d'autres opérateurs. L'acteur doit donc échanger, communiquer avec les autres acteurs de l'activité. Les acteurs ne focalisent pas leur attention sur la communication,celle-ci demeure un outil au service de leurs actions. Au cours d'une activité La communication passe par la transmission de signes, qui, dans le cas d'une communication médiatisée par un langage sont porteurs de significations devant être communes à l'émetteur et au récepteur. La communication sert à la coordination en permettant l'échange d'idées ou de concepts à travers le dialogue et la transmission d'objets intermédiaires. [Michinov 2001] Les objets intermédiaires servent les activités de co-conception et de conception distribuée en assurant une synchronisation cognitive entre les acteurs impliqués.

### 3.1 Le concept objet intermédiaires

La communication au cours de la réalisation d'une activité collective nécessite également la transmission d'**objets intermédiaires** entre les acteurs. La conception d'un objet est « *ponctuée dans le temps* » par la production d'une quantité d'objets intermédiaires comme des idées, des textes, des dessins, des maquettes, etc. Ces objets intermédiaires sont « *des vecteurs de représentation, orientés par une intention ou un objectif issu d'un monde socio-technico-économique lié d'une* façon *ou d'une autre à celui de la réalisation de cet objectif*» [Jeantet et al. 1996]. Ces objets constituent donc la matérialisation des interactions apparaissant entre les acteurs au cours de la conception d'un objet. Les objets intermédiaires participent à l'orientation de l'activité en introduisant des interprétations, des matérialisations d'un état de l'activité en cours de réalisation.

### 3.2 Objets intermédiaires dans la communication

la conception est défini comme l'activité intellectuelle par laquelle sont imaginées quelques dispositions visant à changer une situation existante en une situation préférée [Simon, 1991]. En la contextualisant, elle peut être précisée de la façon suivante «*La conception consiste à donner un ensemble de propositions permettant de décrire le produit (forme, dimension, moyen d'obtention) et répondant globalement à un cahier des charges*» [Tichkiewitch, al 1993]. En effet, elle est située dans le monde des idées, de la connaissance, mais aussi elle relève aussi de la sphère de l'action. Ainsi ne faut-il pas la limiter à une activité intellectuelle : c'est en même temps une activité de création et de décision. En témoigne la production des multiples objets intermédiaires qui la compose, que ceux-ci soient immatériels (règlements, logiciels, modèles numériques) ou matériels (dessins techniques, textes, maquettes). Le but de notre présentation est de montrer que ces objets intermédiaires constituent les vecteurs les plus pertinents des activités de communication omniprésentes dans le processus de conception. Dans la mouvance actuelle qui tend à intégrer de plus en plus ce processus,il devient urgent de prendre en compte la place qu'occupent les objets intermédiaires dans une communication devenue vitale pour la mise en place de la conception intégrée. Cet outil d'analyse doit nous permettre d'entrer au coeur de l'action de concevoir avec une double vue:la première orientée vers le contenu de la conception dont l'objet est la représentation, la seconde orientée vers les interactions entre les acteurs de la conception dont ces objets sont le centre. Ces objets intermédiaires [Mer et Tichkiewitch al. 1995] sont les «*vecteurs les plus pertinents des activités de communication omniprésentes dans le processus de conception* », ils sont à la fois ce qui va définir le produit lui-même et être le support de l'interaction entre les concepteurs ou partenaires.

### 4. Rôle de l'image

Dans le domaine du bâtiment, l'expression des contraintes et la recherche de solution passe prioritairement par des modes d'expression graphique. Pour les architectes, par exemple, les jeux permanents de références et d'analogies conduisent à penser que l'image est un objet intermédiaire qui occupe une place centrale dans leur stratégie de conception [Conan 1990], [Fernandez 2002]. Ainsi, l'image permet de se confronter aux contraintes connues et de faire germer de nouvelles directions pour le projet. L'image est un moyen de communication très important dans le domaine de l'architecture qui intervient dans les différentes phases de la conception d'un projet. Elle peut être considérée comme un outil d'aide à la décision. Les images sont des opérateurs puissants de conversion sensorielle. L'image peut être un support pour la recherche d'informations lors de la phase de conception en Architecture. Au cours d'une recherche d'informations techniques (idée, produit, etc.), l'utilisateur peut acquérir plus rapidement des informations avec des images de produits ou de projets d'architecture qu'il ne le ferait à la lecture de textes [Nakapan et al 2002]. L'image est d'abord un objet auquel on se colle mentalement. C'est seulement dans un second temps qu'on s'en décolle en y appliquant des opérations de transformation. Et c'est alors qu'elle existe comme premier cran pour la pensée. On peut formuler les choses autrement. Les images que nous regardons n'ont pas seulement le pouvoir de donner du sens, mais ont aussi le pouvoir de nous contenir et d'exister comme lieu de transformations multiples.

**4.1 L'image : support communicationnel de l'interaction**

L'image peut être un support pour la recherche d'information lors de la phase de conception en architecture. Au cours d'une recherche d'informations techniques (idée, produit, exemple), l'utilisateur peut acquérir plus rapidement des informations avec des images de produits ou de projet d'architecture qu'il ne le ferait à la lecture de textes. L'images semble un support idéal pour répondre à ces exigence. La lecture d'images est plus rapide et plus conviviale que la lecture du texte. De plus ,l'image représente des informations parfois subjectives à caractère esthétique, qui ne peuvent pas être transcrites sous forme d'un texte. L'interprétation que l'on fait d'une image dépend de la culture et de la langue de la personne qui la regard. Toutefois ,nous pouvons affirmer que « toute image est lisible quelque soit la langue ou la culture de la personne qui l'observe ». Dans toute tâche de conception, la visualisation est une nécessité absolue et concomitante.

Nous baserons notre propos sur l'image en tant que support de cette interaction. Principalement parce que le dessin d'architecte est et reste la trace visible et exploitable du passage de l'idée à la forme ou de la forme à l'idée. Il est étroitement lié à une culture et à une technique de l'architecte. Présent à toutes les étapes de la conception, il est de fait le mode de communication par excellence, même si des explications souvent orales, parfois écrites peuvent l'accompagner.

**4.2 L'usage de l'information véhiculée par les images**

Nous assimilons une information à une collection de données organisées pour donner forme à un message, le plus souvent sous une forme visible, images, écrite, ou orale. La façon d'organiser les donnés résulte d'une intention de l'émetteur et est donc parfaitement subjective.

> *Une connaissance vient s'intégrer dans son système personnel de représentation, pour cela l'information reçue subit une série d'interprétation liées aux croyances générales,au milieu socioprofessionnel,au point de vue à l'intention ,au projet de vue,a l'intention,au projet de l'individu récepteur. Pour qu'une information devienne connaissance, il faut également que le sujet puisse construire une représentation qui fasse sens*

> Contrairement *à l'information la connaissance n'est pas seulement mémoire, item figé dans un stock ; elle reste activable selon une finalité, une intention, un projet* [ Prax 2000]

Le traitement de l'information est à la base de l'IE : c'est-à-dire que de ce processus essentiellement dépend la valeur de l'information pour ses utilisateurs. Traiter l'information; c'est rassembler l'ensemble des données recueillies par les différents canaux pour en faire une synthèse, cohérente et surtout porteuse de sens pour l'utilisateur. Certains modèles d'évaluation des sources et de la valeur de l'information, ainsi que des outils d'aide à la sélection et à la lecture des systèmes d'informations, se sont développés au cours des dernières années. Ils présentent malheureusement le plus souvent des limites, spécifique à certains domaines d'activités.

**5. Perspectives**

Le besoin informationnel d'un utilisateur est un concept qui varie en définition, selon les chercheurs et selon les différents utilisateurs. Il existe des recherches qui ont essayé de lui donner une définition [Campbell et Rijsbergen 1996]. Nous définissons un besoin informationnel comme étant une représentation informationnelle d'un problème décisionnel. Définir un problème décisionnel implique une connaissance sur l'utilisateur et son environnement. Un problème décisionnel peut être considéré comme une fonction d'un modèle de l'utilisateur, de son environnement et de son objectif. De ce fait nous proposons une méthodologie qui prenne en compte l'aspect multi-acteurs à la conception d'un projet. Nous pensons que le besoin informationnel peut être résolu en prenant en compte :

- **L'enjeu** ; c'est ce que l'organisation risque de perdre ou de gagner selon la décision prise.
- **Les caractéristiques individuelles** ; font référence à l'utilisateur (Architecte, Entrepreneur,Client),ses comportements; ses préférences, etc.
- **Les paramètres de l'environnement** ; le cadre dans lequel le projet est établi, les apports de la société sur l'organisation. Ils peuvent être globaux ou immédiats.

Les travaux par [Thiery et David, 2002] sur la personnalisation des réponses en système de recherche d'information ont adapté les quatre phases cognitives identifiées dans le processus d'apprentissage humain, c'est-à-dire :

- **La phase d'observation** : l'apprenant prend connaissance de son environnement par le processus d'observation;
- **La phase d'abstraction élémentaire** : l'apprenant désigne les objets observés par des mots, ce qui correspond également à une phase d'acquisition de vocabulaire;
- **La phase de symbolisation et de raisonnement** : l'apprenant emploie des vocabulaires spécialisés qui relèvent d'un niveau d'abstraction des concepts élevés.

Ce modèle a été transformé en un modèle de l'utilisateur dans un cadre de recherche d'information. Les deux premières phases (observation et abstraction élémentaire) sont combinées dans une seule phase **exploration** qui nous donne un modèle :

M = {Identité, Objectif, {Activité} {Sous sessions}}
Où

*Activité* = {Type-activité, Classification, Evaluation}

*Type* = {Exploration, Requête, Synthèse}

*Classification* = {Attributs, Contraintes}

*Evaluation* = {Solution du système, Degré de pertinence}

- **Identité :** L'identité de l'utilisateur. Ce paramètre permet d'individualiser l'historique des sessions de l'utilisateur.
- **Objectif :** L'objectif principal ou *besoin* de l'utilisateur pour la session.
- **Activité :** Une activité de l'utilisateur pour obtenir des solutions à son besoin en information. Une session est composée de plusieurs activités. Une activité est définie par trois paramètres : *type-activité, classification, évaluation*.
- **Type-activité :** Les types d'activité correspondent aux différentes phases d'habitudes évocatrices, à savoir ici *l'exploration, la* requête *et la synthèse*.
- **Classification :** La classification est l'approche que nous employons pour l'accès à l'information. La technique de classification permet à l'utilisateur d'exprimer sa demande en information dans les phases d'habitudes évocatrices que nous implémentons. L'utilisateur pourra spécifier les *attributs* des documents à classer et les *contraintes* qui doivent être satisfaites par les documents.

**5.1 exemple**

Notre modèle de l'usager est constitué d'un ensemble de triplets : le nom de l'attribut, la valeur de cet attribut et le facteur de certitude sur cette valeur. Lors de la connexion d'un utilisateur, nous initialisons le contenu du modèle usager grâce à des informations héritées du stéréotype auquel l'utilisateur appartient. Pour ce faire nous avons établis un schéma faisant état des différentes étapes qui contribuent à l'élaboration d'un projet en Architecture. Chacune de ces étapes seront prises comme étant de grande famille, à l'intérieur duquel nous établirons une classification de sous éléments qui seront pris à titre d'attribut qui seront associés des valeurs qui correspondent aux besoins des utilisteurs-décideurs. Cette classification est une approche que nous mettons en place pour faciliter l'accès à l'information. En effet, cette technique de classification permettra à l'utilisateurs-décideurs d'exprimer sa demande en information dans les phases d'habitudes évocatives. Exemple dans le cas de la composition d'un dossier de la conception détaillée,nous aurons comme éléments intervenants : dessin technique, plan masse, descriptif du projet, etc. Cette étude est à l'heure actuelle qu'en phase d'élaboration que nous formulons sous forme de balise exemple de la composition d'un dossier:

**{Dossier**
    **{Appel_offre**
        **{Maitre_ouvrage}**
        **{Batiment}**
        **{Image}**
        **}**
    **}**

Un **objet fondation** est décrit par des attributs géométriques, technologiques, technique et économiques ; les attributs techniques et économiques sont instanciers à l'issue des évaluations et vérifications, A partir de ces objets et d'informations plus générales sur l'ouvrage, on modélise le **projet** en utilisant le schéma suivant :

**Projet** = *(identificateur, {sol,...}, {porteur,..}, {fondation,....})*

*{Porteur repose sur fondation,....} et {fondation repose sur sol,...}*

Le modèle conceptuel associé au système d'information du projet est partitionné en objet exemple: parcelle, objets bâtiment et objets réseaux.

**Parcelle** = *{limites latérales/fronts à rue/alignement/axes de circulation/alignements opposés}* ;

**Bâtimen**t= {plates-formes/étages/cellules} ;

## 6. Conclusion

Dans cet article nous avons montrés l'importance des objets intermédiaires (plus particulièrement l'image) comme étant un moyen qui permet la transmission de l'information par les différents partenaires dans le domaine du bâtiment. Ces objets intermédiaires peuvent aider nos acteurs à cerner leurs besoins , de leur permettre la prise de décision dans le cadre de l'élaboration d'une programmation , conception ou réalisation d'un édifice. Notre étude est de voir comment nous pouvons permettre cela ,par le biais d'une étude au préalable l'anticipation des besoins et des enjeux de nos acteurs.